# Updated physics performance of the ESSnuSB experiment

**ESSnuSB collaboration**

A. Alekou[1,2], E. Baussan[3], N. Blaskovic Kraljevic[4], M. Blennow[5,6], M. Bogomilov[7], E. Bouquerel[3], A. Burgman[8], C. J. Carlile[9], J. Cederkall[8], P. Christiansen[8], M. Collins[4,10], E. Cristaldo Morales[11], L. D'Alessi[3,a], H. Danared[4], J. P. A. M. de André[3], J. P. Delahaye[1], M. Dracos[3], I. Efthymiopoulos[1], T. Ekelöf[2], M. Eshraqi[4], G. Fanourakis[12], E. Fernandez-Martinez[13], B. Folsom[4], M. Ghosh[14,15,b], G. Gokbulut[16], L. Halić[14,17], A. Kayis Topaksu[16], B. Kliček[14,c], K. Krhač[14], M. Lindroos[4], M. Mezzetto[18], M. Oglakci[16], T. Ohlsson[5,6], M. Olvegård[2], T. Ota[13], J. Park[8,21], G. Petkov[7], P. Poussot[3], S. Rosauro-Alcaraz[13], G. Stavropoulos[12], M. Stipčević[14], F. Terranova[11], J. Thomas[3], T. Tolba[19], R. Tsenov[7], G. Vankova-Kirilova[7], N. Vassilopoulos[20], E. Wildner[1], J. Wurtz[3], O. Zormpa[12], Y. Zou[2]

[1] CERN, 1211 Geneva 23, Switzerland
[2] Uppsala University, P.O. Box 256, 751 05 Uppsala, Sweden
[3] IPHC, Université de Strasbourg, CNRS/IN2P3, Strasbourg, France
[4] European Spallation Source, Box 176, 221 00 Lund, Sweden
[5] Department of Physics, School of Engineering Sciences, KTH Royal Institute of Technology, Roslagstullsbacken 21, 106 91 Stockholm, Sweden
[6] The Oskar Klein Centre, AlbaNova University Center, Roslagstullsbacken 21, 106 91 Stockholm, Sweden
[7] Faculty of Physics, Sofia University St. Kliment Ohridski, 1164 Sofia, Bulgaria
[8] Department of Physics, Lund University, P.O Box 118, 221 00 Lund, Sweden
[9] Department of Physics and Astronomy, FREIA, Uppsala University, Box 516, 751 20 Uppsala, Sweden
[10] Faculty of Engineering, Lund University, P.O Box 118, 221 00 Lund, Sweden
[11] University of Milano-Bicocca and INFN sez. di Milano-Bicocca, Milan, Italy
[12] Institute of Nuclear and Particle Physics, NCSR Demokritos, Neapoleos 27, 15341 Agia Paraskevi, Greece
[13] Departamento de Fisica Teorica and Instituto de Fisica Teorica, IFT-UAM/CSIC, Universidad Autonoma de Madrid, Cantoblanco, 28049 Madrid, Spain
[14] Center of Excellence for Advanced Materials and Sensing Devices, Ruder Bošković Institute, 10000 Zagreb, Croatia
[15] School of Physics, University of Hyderabad, Hyderabad 500046, India
[16] Department of Physics, Faculty of Science and Letters, University of Cukurova, 01330 Adana, Turkey
[17] Department of Physics, University of Rijeka, 51000 Rijeka, Croatia
[18] INFN sez. di Padova, Padua, Italy
[19] Institute for Experimental Physics, Hamburg University, 22761 Hamburg, Germany
[20] Spallation Neutron Science Center, Dongguan 523803, China
[21] Present address: The center for Exotic Nuclear Studies, Institute for Basic Science, Daejeon 34126, Korea

**Abstract** In this paper, we present the physics performance of the ESSnuSB experiment in the standard three flavor scenario using the updated neutrino flux calculated specifically for the ESSnuSB configuration and updated migration matrices for the far detector. Taking conservative systematic uncertainties corresponding to a normalization error of 5% for signal and 10% for background, we find that there is $10\sigma$ ($13\sigma$) CP violation discovery sensitivity for the baseline option of 540 km (360 km) at $\delta_{CP} = \pm 90°$. The corresponding fraction of $\delta_{CP}$ for which CP violation can be discovered at more than $5\sigma$ is 70%. Regarding CP precision measurements, the $1\sigma$ error associated with $\delta_{CP} = 0°$ is around $5°$ and with $\delta_{CP} = -90°$ is around $14°$ ($7°$) for the baseline option of 540 km (360 km). For hierarchy sensitivity, one can have $3\sigma$ sensitivity for 540 km baseline except $\delta_{CP} = \pm 90°$ and $5\sigma$ sensitivity for 360 km baseline for all values of $\delta_{CP}$. The octant of $\theta_{23}$ can be determined at $3\sigma$ for the values of: $\theta_{23} > 51°$ ($\theta_{23} < 42°$ and $\theta_{23} > 49°$) for baseline of 540 km (360 km). Regarding measurement precision of the atmospheric mixing parameters, the allowed values at $3\sigma$ are: $40° < \theta_{23} < 52°$ ($42° < \theta_{23} < 51.5°$) and $2.485 \times 10^{-3}$ eV$^2 < \Delta m_{31}^2 < 2.545 \times 10^{-3}$ eV$^2$ ($2.49 \times 10^{-3}$

[a] e-mail: loris.dalessi@iphc.cnrs.fr (corresponding author)
[b] e-mail: mghosh@irb.hr (corresponding author)
[c] e-mail: budimir.klicek@irb.hr (corresponding author)

Springer

eV$^2$ < $\Delta m^2_{31}$ < 2.54 × 10$^{-3}$ eV$^2$) for the baseline of 540 km (360 km).

## 1 Introduction

The European Spallation Source neutrino Super-Beam (ESSnuSB) is a proposed accelerator-based long-baseline neutrino experiment in Sweden [1,2]. In this project, high intensity neutrino beam will be produced at the upgraded ESS facility for the ESSnuSB in Lund and these neutrinos will be detected either at the distance of 540 km at Garpenberg mine or at the distance of 360 km at Zinkgruvan mine, both of which are located in Sweden. The primary goal of this experiment is to measure the leptonic CP phase $\delta_{CP}$ by probing the second oscillation maximum. As the variation of neutrino oscillation probability with respect to $\delta_{CP}$ is much higher in the second oscillation maximum as compared to the first oscillation maximum [3–5], ESSnuSB as a second generation super-beam experiment has the potential of measuring $\delta_{CP}$ with unprecedented precision compared to the first generation long-baseline experiments. In the standard three flavor scenario, the phenomenon of neutrino oscillation can be described by three mixing angles: $\theta_{12}$, $\theta_{13}$, and $\theta_{23}$, two mass squared differences $\Delta m^2_{21}$ (= $m^2_2 - m^2_1$), and $\Delta m^2_{31}$ (= $m^2_3 - m^2_1$) and one Dirac type phase $\delta_{CP}$. During the past few decades, some of these parameters are measured with good precision. At the moment, the unknown parameters are: (i) the mass hierarchy of the neutrinos, which can be either normal i.e., $\Delta m^2_{31}$ > 0 or inverted $\Delta m^2_{31}$ < 0, (ii) the octant of the mixing angle $\theta_{23}$, which can be either the lower i.e. $\theta_{23}$ < 45° or the higher i.e., $\theta_{23}$ > 45° and (iii) $\delta_{CP}$. The long-baseline experiments which are currently running to measure these unknowns are T2K [6] in Japan and NO$\nu$A [7] in USA. It is believed that these two experiments will give a hint towards the true nature of the unknown oscillation parameters and the future generation of long-baseline experiments for example ESSnuSB, T2HK [8] and DUNE [9] will establish these facts with significant confidence level. Regarding the true hierarchy of the neutrino mass, the results of both T2K and NO$\nu$A favour normal hierarchy over inverted hierarchy. Regarding the true nature of the octant of $\theta_{23}$ both these experiments support a higher octant, however the maximal value i.e., $\theta_{23}$ = 45° is also allowed within 1$\sigma$. Regarding the value of $\delta_{CP}$, there is a mismatch between T2K and NO$\nu$A. Considering the branch for $\delta_{CP}$ as −180° ≤ $\delta_{CP}$ ≤ 180°, T2K supports the best-fit value of $\delta_{CP}$ around −90° i.e., the maximal CP violating value and the best-fit value measured by NO$\nu$A is around 0° i.e., the CP conserving value. Because of this the best-fit value of $\delta_{CP}$ coming from the global analysis of the world neutrino data is −163° [10]. However it is important to note that both the values of $\delta_{CP}$ = 0° and −90° are allowed at 3$\sigma$ and it requires more data to establish the true value of $\delta_{CP}$.

In this paper we present the physics performance of the ESSnuSB experiment within the standard three flavor scenario for both baseline options of 540 km and 360 km. In particular we will present the capability of the ESSnuSB experiment to measure the current unknowns which were discussed in the previous section. In addition we will present its capability to precisely measure $\Delta m^2_{31}$ and $\theta_{23}$. Note that the physics performance of ESSnuSB within the three flavor scenario has been studied in the past [11–15]. However, in all these studies the configuration of ESSnuSB used was taken from an earlier project. For example, the fluxes and event selection in the form of migration matrices were taken from the MEMPHYS project [16] and the systematics were taken from Ref. [17]. In this paper we will present the updated physics performance of ESSnuSB by considering the new neutrino flux calculated specifically for the ESSnuSB configuration and updated migration matrices for the far detector. The neutrino fluxes used in this work have been calculated by considering a new target and horn focusing, whose design has been optimized by using genetic algorithm calculations [18]. The new design results in an improved statistics compared with the layout of the target station derived from the EUROnu project [1,19,20]. The event selection algorithm used in this work has been optimized for the relatively low neutrino energies of the ESSnuSB beam, increasing the signal selection efficiency from 50% [16] to more than 90%. This resulted in a significant reduction of the statistical error of the experiment. The event selection and reconstruction efficiencies [21] are encoded in the newly produced migration matrices used in this paper.

The paper is organized in the following way. In the next section we will discuss the configuration of the ESSnuSB experiment for which the sensitivities are calculated. In the following section we will present our updated results both in terms of number of events and $\chi^2$. Finally we will summarize our results and conclude on the physics capability of ESSnuSB.

## 2 Experimental and simulation details

For the simulation of the ESSnuSB experiment we consider a water Cherenkov detector of fiducial volume 538 kt located either at a distance of 540 km or 360 km from the neutrino source. Neutrino beam production is driven by a powerful linear accelerator (linac) capable of delivering 2.7 × 10$^{23}$ protons on target per year having a beam power of 5 MW with proton kinetic energy of 2.5 GeV. The fluxes [18] and the event selection [21] for the Far Detectors are calculated using full Monte Carlo simulations specific to the ESSnuSB conditions. Neutrino interactions are modelled using GENIE 3.0.6 neutrino interaction generator [22–24]. The detector response and efficiencies are calculated using the same detector parameters as for the Hyper-K detector [8] with 40% photomultiplier (PMT) coverage, while the expected event rate is scaled to 538 kt fiducial mass foreseen by the ESSnuSB project. Particle propagation and detector response are simulated using a GEANT4-based [25–27] software named WCSim [28], specifically for designing water-based Cherenkov detectors. The event selection and charged particle momentum reconstruction are based on the fiTQun reconstruction software [29,30]. Since the dimensions and PMT coverage of Hyper-K detector are very similar to the ESSnuSB design, we do not expect a significant difference in detector response. Full simulation of the ESSnuSB specific detector is currently under production. These fluxes and detector response with efficiencies encoded by migration matrices are then incorporated in GLoBES [31,32] to calculate event rates and $\chi^2$. We have checked that the event rates obtained by Monte Carlo and the event rates generated by GLoBES are consistent. We have considered a conservative estimate of the systematic errors on the overall normalization of the expected number of detected events at the Far Detectors: 5% for signal and 10% for background, unless otherwise specified. No systematic effects on the shape of the detected energy spectrum have been implemented. The systematic errors are set to be the same for appearance and disappearance channels and also for neutrinos and antineutrinos. We have considered a total run-time of 10 years divided into 5 years of neutrino beam and 5 years of antineutrino beam, unless otherwise specified. The configurations mentioned above are the same for both baseline options of ESSnuSB.

## 3 Results

In this section we present the physics sensitivities of the ESSnuSB experiment. First we will present a discussion on the appearance probability level to understand the energy spectrum to which ESSnuSB is sensitive to. Then we will study the total number of expected events and event spectrum of ESSnuSB. Finally, we will discuss the physics sensitivity of this experiment with respect to the current unknowns in the standard three flavor neutrino oscillation scenario. For the estimation of the sensitivity we calculate the statistical $\chi^2$ using the following formula:

$$\chi^2_{\text{stat}} = 2 \sum_{i=1}^{n} \left[ N_i^{\text{test}} - N_i^{\text{true}} - N_i^{\text{true}} \log \left( \frac{N_i^{\text{test}}}{N_i^{\text{true}}} \right) \right], \quad (1)$$

where $N^{\text{test}}$ is the number of events in the test spectrum, $N^{\text{true}}$ is the number of events in the true spectrum and $i$ is the number of energy bins. The systematic uncertainties are incorporated by the method of pull [33,34]. Unless otherwise specified, the best-fit values of the oscillation parameters are adopted from NuFIT [10] and we list them in Table 1. We present all our results for the normal hierarchy of the neutrino masses.

**Table 1** The best-fit value of the oscillation parameters used in our calculation as given in Ref. [10]

| Parameter | Best-fit value |
|---|---|
| $\theta_{12}$ | 33.44° |
| $\theta_{13}$ | 8.57° |
| $\theta_{23}$ | 49.2° |
| $\delta_{\text{CP}}$ | −163° |
| $\Delta m^2_{21}$ | $7.42 \times 10^{-5}$ eV$^2$ |
| $\Delta m^2_{31}$ | $2.517 \times 10^{-3}$ eV$^2$ |

### 3.1 Discussion at the probability level

As the sensitivity to $\delta_{\text{CP}}$ mainly comes from the appearance channel ($\nu_\mu \to \nu_e$), we plotted only the appearance probability and flux × cross-section vs energy in Fig. 1.

The left panel is for neutrinos and the right panel is for antineutrinos. In each panel the purple curve corresponds to the ESSnuSB baseline option of 540 km and the red curve corresponds to the ESSnuSB baseline option of 360 km. The black dotted curve corresponds to the muon neutrino flux × cross-section. The energy region covered by the black dotted curve shows the energy spectrum to which ESSnuSB is sensitive. For purple and red curves, the solid line corresponds to the value of $\delta_{\text{CP}} = -90°$ and the dashed line corresponds to the value of $\delta_{\text{CP}} = 0°$. Values used for oscillation parameters other than $\delta_{\text{CP}}$ are given in Table 1. We note that ESSnuSB is sensitive to the second oscillation maximum for the baseline option of 540 km, while it is sensitive to some part of the first oscillation maximum and some part of the second oscillation maximum for the baseline option of 360 km. However, for the negative polarity, the ESSnuSB baseline option of 540 km is also sensitive to the third oscillation maximum. We also note that for a given color, the separation in height between the solid curve and dashed curve are more pronounced in the second oscillation maximum as compared to the first oscillation maximum. This shows the fact that the variation of oscillation probability with respect to $\delta_{\text{CP}}$ is much more around the second oscillation maximum as compared to the first oscillation maximum. Therefore we expect an unprecedented CP sensitivity of ESSnuSB.

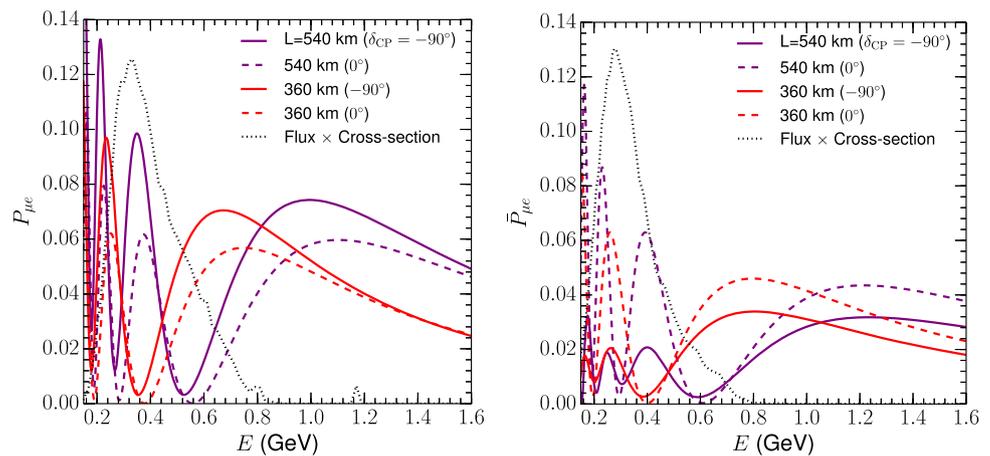

**Fig. 1** Appearance channel probability and flux × cross-section vs energy. The left panel is for neutrinos and the right panel is for antineutrinos

**Table 2** Signal and background events for the appearance channel corresponding to positive (negative) polarity per year

|  | Channel | $L = 540$ km | $L = 360$ km |
|---|---|---|---|
| Signal | $\nu_\mu \to \nu_e$ ($\bar{\nu}_\mu \to \bar{\nu}_e$) | 292.85 (70.04) | 557.83 (118.80) |
| Background | $\nu_\mu \to \nu_\mu$ ($\bar{\nu}_\mu \to \bar{\nu}_\mu$) | 20.41 (4.41) | 68.15 (13.81) |
|  | $\nu_e \to \nu_e$ ($\bar{\nu}_e \to \bar{\nu}_e$) | 133.24 (25.13) | 298.70 (57.14) |
|  | $\bar{\nu}_e \to \bar{\nu}_e$ ($\nu_e \to \nu_e$) | 0.08 (0.92) | 0.20 (2.11) |
|  | $\nu_\mu$ NC ($\bar{\nu}_\mu$ NC) | 14.16 (2.27) | 31.86 (5.11) |
|  | $\bar{\nu}_\mu \to \bar{\nu}_e$ ($\nu_\mu \to \nu_e$) | 2.31 (5.63) | 3.99 (11.70) |
|  | $\nu_e \to \nu_\mu$ ($\bar{\nu}_e \to \bar{\nu}_\mu$) | 0.04 (–) | 0.08 (–) |
|  | $\bar{\nu}_\mu \to \bar{\nu}_\mu$ ($\nu_\mu \to \nu_\mu$) | 0.14 (0.49) | 0.45 (1.26) |
|  | $\bar{\nu}_\mu$ NC ($\nu_\mu$ NC) | 0.24 (0.43) | 0.54 (0.96) |
|  | $\nu_e$ NC ($\bar{\nu}_e$ NC) | 0.57 (–) | 1.29 (–) |

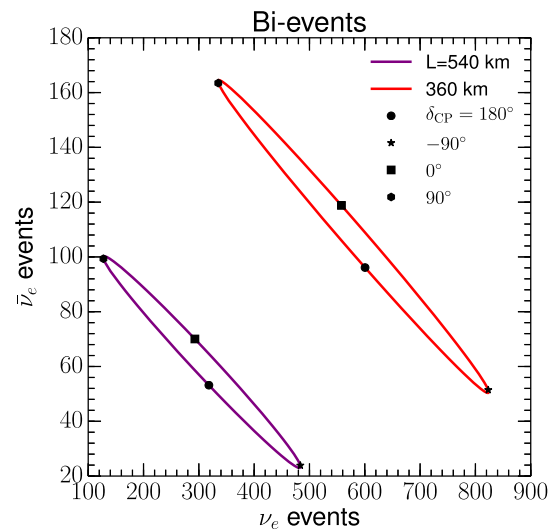

**Fig. 2** The bi-event distribution for both baselines in the $\nu_e$ events vs $\bar{\nu}_e$ events plane. Different values of $\delta_{CP}$ are shown by black markers

### 3.2 Discussion at the event level

In this section we present the total event rates and event spectrum of ESSnuSB for both appearance and disappearance channels ($\nu_\mu \to \nu_\mu$), for both positive and negative polarities and for both baseline options of ESSnuSB. The oscillation parameters which are used in these calculations are as given in Table 1, except the value of $\delta_{CP}$. For $\delta_{CP}$, we have taken the value as $0°$. All the numbers are generated for one year running of ESSnuSB.

In Table 2, we present the total number of the appearance channel events for signal and background which were considered in our analysis.

The sensitivity to mass hierarchy, octant of $\theta_{23}$ and $\delta_{CP}$ comes from the appearance channel. From this table we notice that for both baseline options of ESSnuSB, the number of events for the positive polarity is higher as compared to the number of events in the negative polarity. The reason is that for a given run-time, both the neutrino fluxes and neutrino cross-sections are higher than the antineutrino fluxes and antineutrino cross-sections. Further we notice that the number of events for the ESSnuSB baseline option of 360 km is much higher as compared to the ESSnuSB baseline option of 540 km for both polarities. The reason is that as the baseline $L$ increases, the flux falls as $1/L^2$. Therefore we expect that the physics performance of the 360 km baseline option of ESSnuSB will be better than the physics performance of the 540 km baseline option because of higher statistics. The major backgrounds in the appearance channel for positive polarity are: the intrinsic $\nu_e$ beam component, the misidentified $\nu_\mu \to \nu_\mu$ events, neutral current, and wrong sign backgrounds i.e., $\bar{\nu}_\mu \to \bar{\nu}_e$. Similarly, for negative polarity these are: $\bar{\nu}_e$ beam, $\bar{\nu}_\mu \to \bar{\nu}_\mu$, neutral current and $\nu_\mu \to \nu_e$.

In Fig. 2, we present the bi-event plot i.e., total $\nu_e$ events on the x-axis and $\bar{\nu}_e$ events on the y-axis. It is well known that in the $\nu_e$–$\bar{\nu}_e$ plane, different values of $\delta_{CP}$ form an ellipse [35]. In this panel, the purple ellipse corresponds to the ESSnuSB baseline option of 540 km and the red ellipse corresponds to the baseline option of 360 km. The number of events corresponding to different values of $\delta_{CP}$ are shown by black markers. This figure shows the variation of events in the appearance channel as $\delta_{CP}$ varies between different values. As this variation is quite large, we expect good CP sensitivity for ESSnuSB. We also notice that the red ellipse is stretched more in both x-axis and y-axis, as compared to the purple ellipse. Therefore the CP sensitivity of ESSnuSB will be higher for the baseline option of 360 km as compared to the baseline option of 540 km. In Fig. 3 we plot the event spectrum corresponding to signal and major backgrounds for the appearance channel as a function of reconstructed energy. The top row is for the baseline option of 540 km and the bottom row is for the baseline option of 360 km. In each row, the left panel is for positive polarity and the right panel is for negative polarity. In all panels we notice that in the energy region where the signal peaks, the major contribution for the backgrounds comes from the $\nu_e/\bar{\nu}_e$ beam for positive/negative polarity.

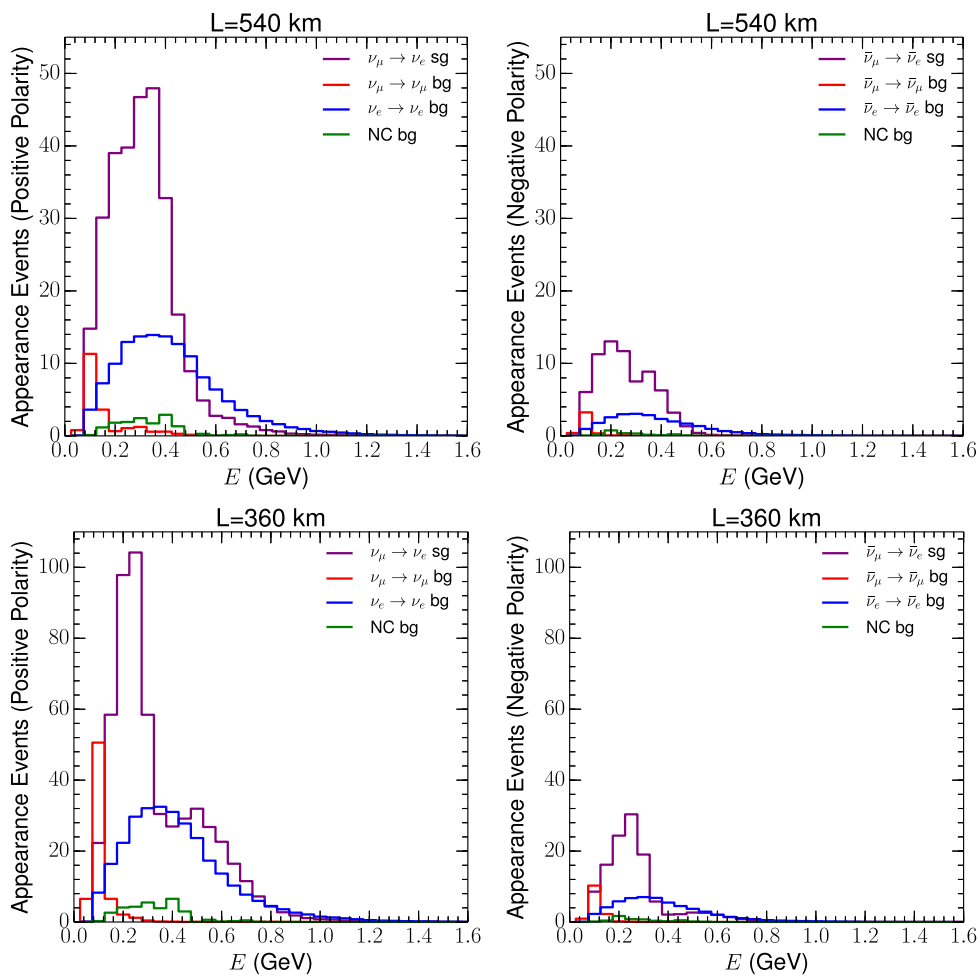

**Fig. 3** Appearance channel event spectrum vs reconstructed energy. The upper panels are for the baseline option of 540 km and the lower panels are for the baseline option of 360 km. Note the difference in scales between upper and lower panels

In Table 3, we present the total number of the disappearance channel events for signal and background which were considered in our analysis.

**Table 3** Events for signal and background corresponding to positive (negative) polarity for 1 year

|  | Channel | $L = 540$ km | $L = 360$ km |
|---|---|---|---|
| Signal | $\nu_\mu \to \nu_\mu$ ($\bar{\nu}_\mu \to \bar{\nu}_\mu$) | 3078.8 (603.53) | 7119.28 (1481.52) |
| Background | $\nu_e \to \nu_e$ ($\bar{\nu}_e \to \bar{\nu}_e$) | 13.44 (0.07) | 29.50 (0.16) |
|  | $\nu_\mu$ NC ($\bar{\nu}_\mu$ NC) | 38.48 (5.92) | 86.59 (13.32) |
|  | $\nu_\mu \to \nu_e$ ($\bar{\nu}_\mu \to \bar{\nu}_e$) | 11.67 (0.031) | 35.66 (0.07) |
|  | $\nu_e \to \nu_\mu$ ($\bar{\nu}_e \to \bar{\nu}_\mu$) | 2.86 (0.62) | 7.48 (1.17) |
|  | $\bar{\nu}_\mu \to \bar{\nu}_\mu$ ($\nu_\mu \to \nu_\mu$) | 25.42 (67.86) | 52.21 (131.07) |
|  | $\nu_e$ NC ($\bar{\nu}_e$ NC) | 0.57 (0.10) | 1.29 (0.23) |
|  | $\bar{\nu}_\mu$ NC ($\nu_\mu$ NC) | 0.50 (1.06) | 1.12 (2.37) |
|  | $\bar{\nu}_\mu \to \bar{\nu}_e$ ($\nu_\mu \to \nu_e$) | – (0.30) | – (1.07) |
|  | $\bar{\nu}_e \to \bar{\nu}_e$ ($\nu_e \to \nu_e$) | – (0.12) | – (0.28) |

The disappearance channel contributes mainly in the precision measurement of $\theta_{23}$ and $\Delta m^2_{31}$. For the disappearance channel as well, more events are expected for the baseline option of 360 km and for the positive polarity. The major sources of background contributing to the disappearance channel are $\nu_e \to \nu_e$, neutral current, $\nu_\mu \to \nu_e$ and $\bar{\nu}_\mu \to \bar{\nu}_\mu$ for the positive polarity and $\bar{\nu}_e \to \bar{\nu}_e$, neutral current, $\bar{\nu}_\mu \to \bar{\nu}_e$ and $\nu_\mu \to \nu_\mu$ for the negative polarity. We plot the event spectrum corresponding to the signal and major backgrounds for the disappearance channel as a function of energy in Fig. 4.

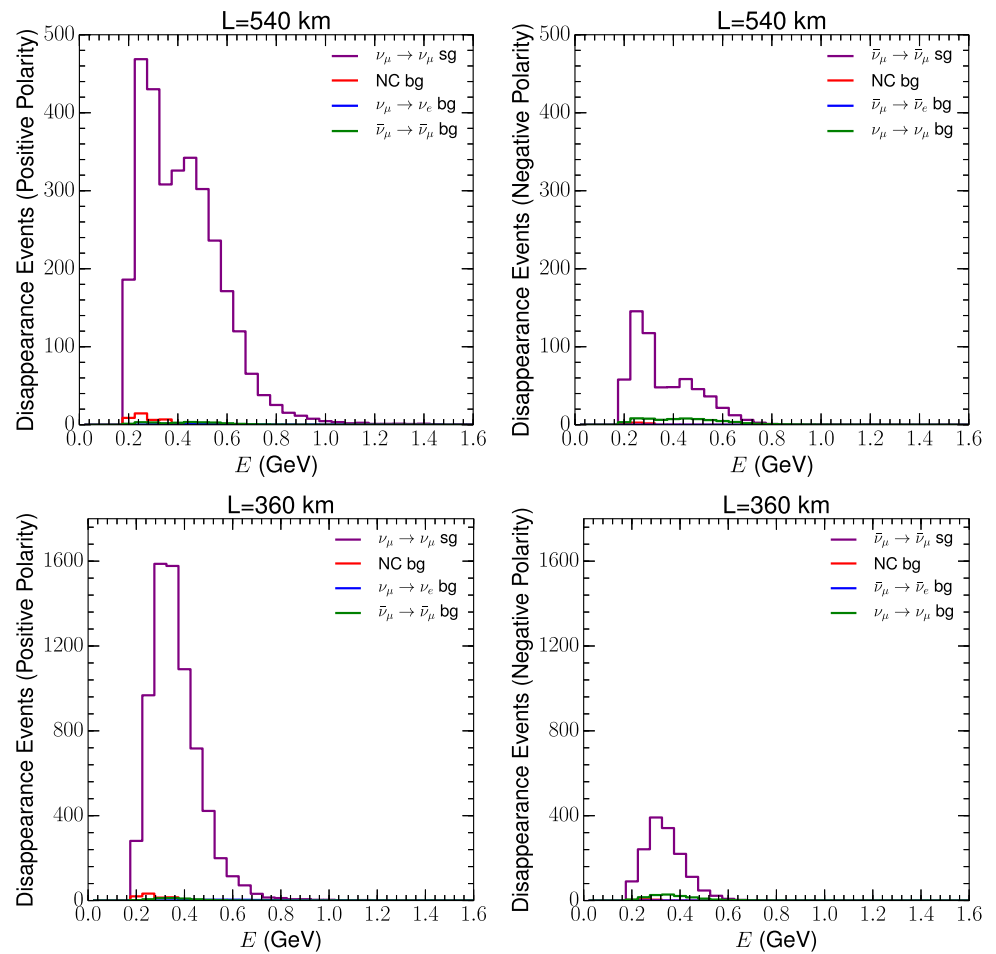

**Fig. 4** Disappearance channel event spectrum vs reconstructed energy. The upper panels are for the baseline option of 540 km and the lower panels are for the baseline option of 360 km. Note the difference in scales between upper and lower panels

The top row is for the baseline option of 540 km and the bottom row is for the baseline option of 360 km. In each row, the left panel is for positive polarity and the right panel is for negative polarity. From the plots we can see that the contribution of the backgrounds is very small for the disappearance channel.

### 3.3 Sensitivity to the unknown parameters

Now we will discuss the capability of the ESSnuSB experiment to measure the current unknowns in the standard three flavor scenario. In Fig. 5, we present the CP violation discovery potential of ESSnuSB for both baseline options. The CP violation discovery potential of an experiment is defined by its capability to distinguish a value of $\delta_{CP}$ other than $0°$ and $180°$. In these panels we use the true values of the parameters as defined in Table 1 and in the test spectrum we have minimized over the neutrino mass hierarchy and $\theta_{23}$ in the range between $40°$ and $52°$. In all the panels, the purple curve corresponds to the baseline option of 540 km and the red curve corresponds to the baseline option of 360 km. In the top left panel we present the CP violation discovery sensitivity as a function of $\delta_{CP}$ (true). From this panel we note that for maximal values of $\delta_{CP}$ around $\pm 90°$, the sensitivity is ca $10\sigma$ for the baseline option of 540 km and ca $13\sigma$ for the baseline option of 360 km. In the top right panel we have plotted the fraction of $\delta_{CP}$ values for which CP violation can be discovered at more than $5\sigma$ as a function of run-time. A run-time of $t$ implies running $t/2$ years in neutrino mode and running $t/2$ years in antineutrino mode. The black horizontal lines correspond to the benchmark of 50% and 70% CP coverage for which CP violation can be discovered at more than $5\sigma$ respectively. From this panel we note that for a nominal running time of two years, we can have $5\sigma$ coverage for 50% values of $\delta_{CP}$. The range expands to 70% values of $\delta_{CP}$ for a running time of 10 years for both baseline options. If we continue to run the experiment for 20 years, then we can have a coverage of around 80%. In the bottom left panel we present the CP violation discovery sensitivity for $\delta_{CP} = -90°$ which is the current best-fit value as obtained from the T2K experiment as a function of run-time. From this panel we can see that for a nominal running time of two years, the sensitivity is always higher than $5\sigma$ and for 20 years of running it goes up to $13\sigma$ for the baseline option of 540 km and $16\sigma$ for the baseline option of 360 km. Finally in the right panel we plot the CP violation discovery sensitivity for $\delta_{CP} = -90°$ as a function of systematic errors assuming the event statistics to be that of 10 years data collection. A value of $x$ in the x-axis implies a systematic error of $x\%$ in the signal and an error of $2x\%$ in the background. From this panel we can see that for the most optimistic set of systematic errors, i.e., 1% error in signal and 2% error in background, we can have around $17\sigma$ sensitivity for the baseline option of 540 km and $20\sigma$ sensitivity for the baseline option of 360 km. However, when we increase the systematics to the most conservative set, i.e., an error of 10% in signal and 20% in background, the sensitivity reaches $8\sigma$ for the baseline option of 540 km and $9.5\sigma$ for the baseline option of 360 km. In both of the bottom panels, the black horizontal lines correspond to the benchmark of $5\sigma$ and $10\sigma$ sensitivity, respectively. From all these four panels we note that the sensitivity for the baseline option of 360 km is superior to the sensitivity of the 540 km baseline.

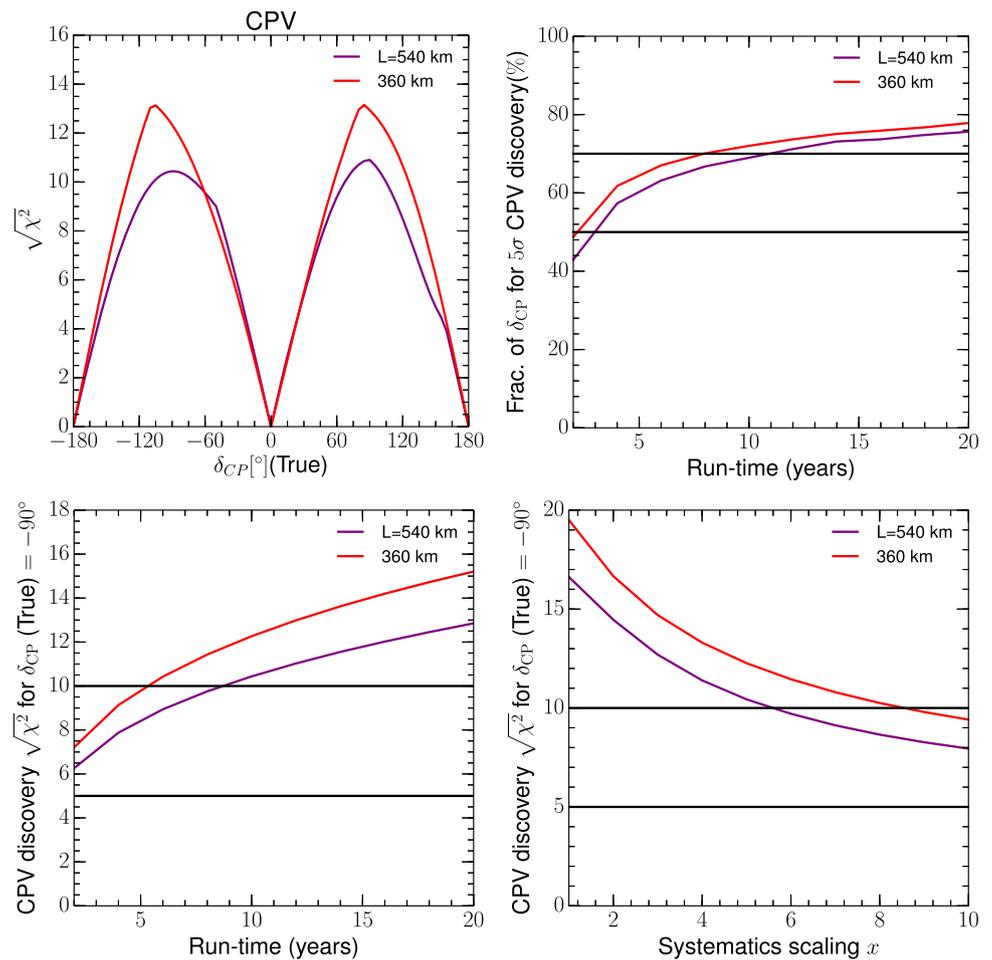

**Fig. 5** CP violation discovery sensitivity of ESSnuSB. The top left panel shows the sensitivity as a function of true $\delta_{CP}$. The top right panel shows the fraction of true values of $\delta_{CP}$ for which CP violation can be discovered at $5\sigma$ as a function of run-time. The left bottom panel shows the sensitivity corresponding to $\delta_{CP} = -90°$ as a function of run-time and the right panel shows the dependence of the sensitivity on the systematics uncertainties $x\%$ on signal and $2x\%$ on background assuming 10 years of data collection

In Fig. 6, we present the CP precision capability of ESS-nuSB. The CP precision capability of an experiment is defined by its potential to distinguish a true value of $\delta_{CP}$ from any other value of $\delta_{CP}$. In these panels we also use the true values of the parameters as defined in Table 1. In the test spectrum we have minimized over the neutrino mass hierarchy and $\theta_{23}$ in the range between $40°$ and $52°$. In the left panel we have plotted the $1\sigma$ error in the measurement of $\delta_{CP}$ as a function of $\delta_{CP}$ (true). The purple curve is for the baseline option of 540 km and the red curve is for the baseline option of 360 km. From this panel we note that the error associated with $\delta_{CP}$ is around $5°$ if the true values of $\delta_{CP}$ are around $0°$ or $180°$ for both baseline options. However, for $\delta_{CP} = -90°$, the error is around $14°$ for the baseline option of 540 km and only $7°$ for the baseline option of 360 km. In the middle and right panels we present the CP precision in the true $\delta_{CP}$ vs test $\delta_{CP}$ plane. The middle panel is for the baseline option of 540 km and the right panel is for the baseline option of 360 km. In each panel, the purple/red/blue curve corresponds to the $1\sigma/2\sigma/3\sigma$ contours, respectively. In an ideal situation,

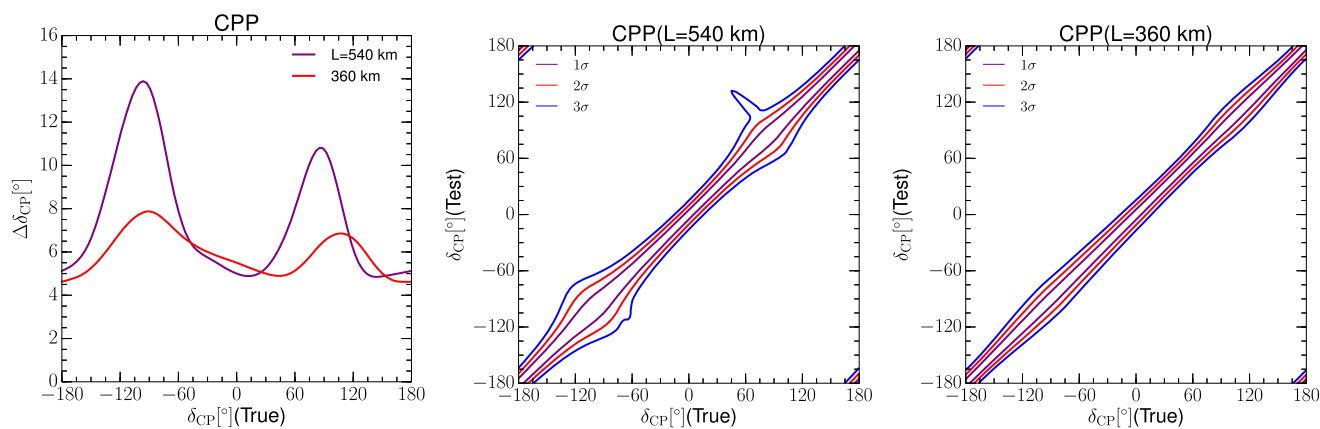

**Fig. 6** CP precision sensitivity of ESSnuSB. Left panel shows the $1\sigma$ error associated with a value of $\delta_{CP}$ as a function of $\delta_{CP}$ (true). The middle and right panels depict the CP precision in the $\delta_{CP}$ (true) vs $\delta_{CP}$ (test) plane

we expect a straight line corresponding to $\delta_{CP}$ (true) = $\delta_{CP}$ (test). Therefore the width of the contours represents the error associated at that given C.L. From these panels we note that the precision at $\delta_{CP} = \pm 90°$ is worse than the precision at $\delta_{CP} = 0°$ and $180°$ [36]. In the middle panel we notice an extended region around $\delta_{CP} = 90°$ for the $3\sigma$ contour. This occurs due to the hierarchy - $\delta_{CP}$ degeneracy [36–38]. From these panels we see again that for $\delta_{CP} = \pm 90°$, the CP precision is better for the baseline option of 360 km, while for $\delta_{CP} = 0°$ and $180°$ the CP precision is very similar in the two cases.

Comparing the results shown in Figs. 5 and 6 with other next-generation long-baseline neutrino experiments [8,9], one can see that ESSnuSB is expected to perform significantly better. This is for the following reasons: (i) the neutrino production will be driven by the powerful 5 MW ESS linac, which will produce the most intense neutrino flux to date, allowing a significant statistical sample to be collected at the second oscillation maximum; (ii) the unique feature of this experiment to probe the second oscillation maximum, where the variation of the appearance channel probability with respect to $\delta_{CP}$ is close to three times higher than that of the first oscillation maximum, making the experiment more resilient to systematic uncertainties; and (iii) the lower neutrino energy implies a smaller rate of non-quasielastic neutrino scattering events, which allows the experiment to obtain a rather pure appearance signal sample while retaining an overall event selection efficiency of higher than 90%.

Our results have been obtained assuming a conservative (5% signal, 10% background) systematic uncertainty on the normalization of signal and background spectra without taking into account the uncertainty on their shapes. The expected sensitivity is quite robust w.r.t the exact assumed value of these uncertainties, as shown in the lower right panel in Fig. 5. To go further, we are currently studying the effects of spectral shape uncertainty. The preliminary results show that there is no significant degradation of sensitivity up to 10% bin-to-bin uncorrelated error. This may be explained by the fact that our measurements can be well approximated as a counting experiment at the second oscillation maximum in which the shape information enters only as a second order effect.

In Fig. 7, we present the hierarchy and octant sensitivity of ESSnuSB. In the left panel we present the hierarchy sensitivity as a function of $\delta_{CP}$ (true). The hierarchy sensitivity of an experiment is defined as its capability to exclude the wrong neutrino mass hierarchy. In this panel we use true values of the parameters as defined in Table 1. In the test spectrum we have minimized over $\theta_{23}$ in the range between $40°$ and $52°$. The purple curve corresponds to the baseline option of 540 km and the red curve corresponds to the baseline option of 360 km. The black horizontal lines correspond to the benchmark of $3\sigma$ and $5\sigma$ sensitivity, respectively. From this panel we understand that for the baseline option of 540 km, one can have a $3\sigma$ hierarchy sensitivity except for $\delta_{CP} = \pm 90°$, and for the baseline option of 360 km one can have a hierarchy sensitivity of $5\sigma$ for all the values of $\delta_{CP}$. From this panel it is evident that the hierarchy sensitivity for the baseline option of 360 km is much better as compared to the baseline option of 540 km. This is because the hierarchy sensitivity depends on the matter effect. Higher matter effect implies higher hierarchy sensitivity. Further, the matter term in the oscillation probability depends on the energy of the neutrinos [39]. As the matter effect is more significant near the first oscillation maximum due to the higher energy, the baseline option of 360 km provides better hierarchy sensitivity as compared to the baseline option of 540 km.

In the middle and left panels we present the octant sensitivity in the $\theta_{23}$ (true) vs $\delta_{CP}$ (true) plane. The octant sensitivity of an experiment is defined by its capability to exclude the wrong octant of $\theta_{23}$. In these panels, we use true values of the parameters as defined in Table 1. In the test spectrum we have minimized over the neutrino mass hierarchy. The mid-dle panel is for the baseline option of 540 km and the right panel is for the baseline option of 360 km. In each panel the purple/red/blue curve corresponds to the $1\sigma/2\sigma/3\sigma$ contours, respectively. The values of $\theta_{23}$ which are plotted in the x-axis, correspond to the current allowed $3\sigma$ values of $\theta_{23}$. In these panels, the region around $\theta_{23} = 45°$ shows the values of $\theta_{23}$ for which the octant cannot be determined at that given C.L. From these panels we see that the octant sensitivity of ESSnuSB is limited. For the baseline option of 540 km, the octant can be determined at $3\sigma$ only if $\theta_{23}$ is greater than $51°$. For the baseline option of 360 km, the octant can be determined at $3\sigma$ except for the $\theta_{23}$ values of $42° < \theta_{23} < 49°$. Clearly, the octant sensitivity for the baseline option of 360 km is slightly better as compared to the octant sensitivity for the baseline option of 540 km.

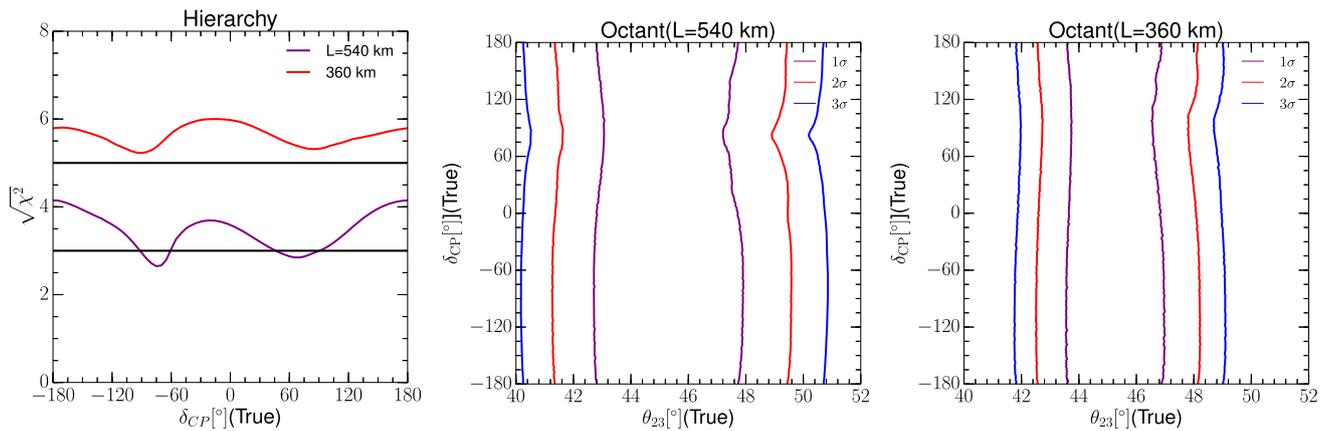

**Fig. 7** Hierarchy and octant sensitivity of ESSnuSB. The left panel corresponds to the hierarchy sensitivity as a function of $\delta_{CP}$ (true). The middle and right panels correspond to the octant sensitivity in the $\theta_{23}$ (true) - $\delta_{CP}$ (true) plane

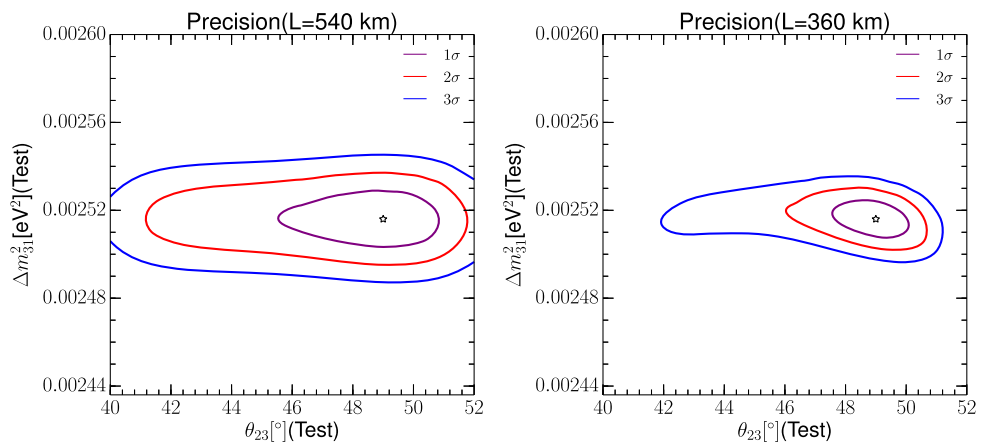

**Fig. 8** Sensitivity to the precision measurement of the atmospheric mixing parameters $\theta_{23}$ - $\Delta m^2_{31}$. The left and right panels are for the baseline options of 540 km and 360 km respectively

Finally, in Fig. 8, we plot the precision measurement of the atmospheric mixing parameters of ESSnuSB in the $\theta_{23}$ (test) - $\Delta m^2_{31}$ (test) plane. In these panels we use the true values of the parameters as defined in Table 1, except for $\delta_{CP}$. For $\delta_{CP}$ we have taken the value as $-90°$ which is the current best-fit value from T2K. The left panel is for the baseline option of 540 km and the right panel is for the baseline option of 360 km. In each panel, the purple/red/blue curve corresponds to the $1\sigma/2\sigma/3\sigma$ CL contours, respectively. The ranges of $\theta_{23}$ and $\Delta m^2_{31}$ axes are the current allowed $3\sigma$ values of these parameters according to the experimental data stored in NuFIT [10]. The measured central values of $\theta_{23}$ and $\Delta m^2_{31}$ are indicated by a star. From these panels we understand that the capability of ESSnuSB to constrain $\Delta m^2_{31}$ is quite good, while its capability to constrain $\theta_{23}$ is limited. This is partially because of the limited octant capability of this experiment. The present best-fit value of $\theta_{23}$ is in the higher octant, and due to the limited octant sensitivity, the region in the lower octant is allowed. For the baseline option of 540 km, all the values of $\theta_{23}$ are allowed at $3\sigma$ and the allowed values of $\Delta m^2_{31}$ are $2.485 \times 10^{-3}$ eV$^2$ to $2.545 \times 10^{-3}$ eV$^2$ at $3\sigma$. For the baseline option of 360 km, the allowed values are $42° < \theta_{23} < 51.5°$ and $2.49 \times 10^{-3}$ eV$^2 < \Delta m^2_{31} < 2.54 \times 10^{-3}$ eV$^2$. In terms of the precision of the atmospheric mixing parameters, the capability of the 360 km baseline is significantly better than the 540 km baseline.

## 4 Summary and conclusion

ESSnuSB is a forthcoming accelerator-based long-baseline neutrino oscillation experiment to be located in Sweden. The primary goal of this experiment is to measure the leptonic CP phase $\delta_{CP}$ at high precision by probing the phenomenon of neutrino oscillations at the second oscillation maximum. In this paper we have studied the physics performance of this experiment in the standard three flavor framework. In particular, we have studied the capability of this experiment to measure the current unknowns in the oscillation parameters which are: neutrino mass hierarchy, octant of atmospheric mixing angle $\theta_{23}$, the leptonic phase $\delta_{CP}$, and the precision of the atmospheric mixing parameters $\theta_{23}$ and $\Delta m_{31}^2$. The physics performance of the ESSnuSB experiment has been studied in the past using the configuration of the MEMPHYS project. In this paper, we have taken the new neutrino flux calculated specifically for the ESSnuSB configuration and updated migration matrices for the far detector. The neutrino fluxes used in this work have been calculated by considering a new target and horn focusing, whose design has been optimized by using genetic algorithm calculations [18]. The new design results in an improved statistics compared with the layout of the target station derived from the EUROnu project [1,19,20]. The event selection algorithm has been optimized for the relatively low neutrino energies of the ESSnuSB beam, increasing the signal selection efficiency from 50% [16] to more than 90%, which was encoded in the new set of migration matrices. At the probability level, we have shown that the variation of the appearance channel probability with respect to $\delta_{CP}$ is large at the second oscillation maximum as compared to the first oscillation maximum. ESSnuSB will therefore have an unprecedented precision of $\delta_{CP}$ measurement. We also have shown that the baseline option of 540 km mainly covers the second oscillation maximum, whereas the baseline option of 360 km covers both the first and second oscillation maxima. At the event level, we have shown that the number of events at the 360 km baseline is larger than the 540 km one because of the shorter baseline. Therefore we expect the sensitivity for the 360 km baseline will be better than that for the 540 km baseline. In this context we also discussed the major background which can affect the sensitivity. Taking an overall conservative systematic normalization error of 5% for signal and 10% for background, we have shown that the CP violation discovery sensitivity is $10\sigma$ ($13\sigma$) for the baseline option of 540 km (360 km) at $\delta_{CP} = \pm 90°$. The corresponding fraction of $\delta_{CP}$ for which CP can be discovered at more than $5\sigma$ is 70%. We have further shown that the CP violation discovery sensitivity is always larger than $5\sigma$ for $\delta_{CP} = -90°$ and the CP coverage at $5\sigma$ is around 44% (50%) even for a nominal run of 2 years for the 540 km (360 km) baseline. This increases to around $13\sigma$ ($15\sigma$) and 76% (80%) respectively when the run-time is increased to 20 years for the baseline option of 540 km (360 km). Then we have also checked how the sensitivity varies when the systematic uncertainty is varied. We have found that even for large systematic errors of 10% signal and 20% background, the CP violation discovery sensitivity is always greater than $5\sigma$ for $\delta_{CP} = -90°$ in 10 years. Regarding CP precision, the $1\sigma$ error associated with $\delta_{CP} = 0°$ is around $5°$ for both of the baseline options and the error associated with $\delta_{CP} = -90°$ is around $14°$ ($7°$) for the baseline option of 540 km (360 km). For neutrino mass hierarchy, one can achieve $3\sigma$ sensitivity for the 540 km baseline except for the true values of $\delta_{CP} = \pm 90°$ and $5\sigma$ sensitivity for the 360 km baseline for all values of $\delta_{CP}$. The values of $\theta_{23}$ for which the octant can be determined at $3\sigma$ is $\theta_{23} > 51°$ ($\theta_{23} < 42°$ and $\theta_{23} > 49°$) for the baseline of 540 km (360 km). Regarding the precision of the atmospheric mixing parameters, the allowed values at $3\sigma$ are: $40° < \theta_{23} < 52°$ ($42° < \theta_{23} < 51.5°$) and $2.485 \times 10^{-3}$ eV$^2 < \Delta m_{31}^2 < 2.545 \times 10^{-3}$ eV$^2$ ($2.49 \times 10^{-3}$ eV$^2 < \Delta m_{31}^2 < 2.54 \times 10^{-3}$ eV$^2$) for the baseline of 540 km (360 km). To summarise, ESSnuSB is a powerful experiment to measure $\delta_{CP}$ with an unprecedented precision compared with all currently planned long-baseline experiments. This experiment also provides the possibility to measure the hierarchy and $\Delta m_{31}^2$ with good precision. Among the two baseline options, 360 km provides the better sensitivity.

Note that the results presented in this work are provisional since the implementation of the systematics is simplistic and the detector response has been determined using the Hyper-K geometry. In the final analysis we will incorporate the near detector which will enable us to implement a more realistic treatment of systematics. This will include correlated systematics between the far and the near detectors, bin-to-bin correlations and shape uncertainties among the other improvements. The full simulation of the ESSnuSB Far Detector response using their exact geometry is currently underway, which will result in an updated migration matrices. We do not expect them to differ much since the foreseen geometry of the ESSnuSB far detector tank does not differ much with respect to that of Hyper-K. Further, as the far detector of this experiment will be underground, there is also the possibility of including the atmospheric data sample in the analysis. This will further improve the hierarchy sensitivity, octant sensitivity and precision sensitivity of the atmospheric mixing parameters.

**Acknowledgements** This project received funding from the European Union's Horizon 2020 research and innovation programme under grant agreement No. 777419. This work has been in part funded by the Deutsche Forschungsgemeinschaft (DFG, German Research Foundation)-Projektnummer 423761110. This work has been in part funded by Ministry of Science and Education of Republic of Croatia grant No. KK.01.1.1.01.0001. T.O. acknowledges support by the Swedish Research Council (Vetenskapsrådet) through Contract No. 2017-03934. We want to thank Cristovao Vilela, Erin O'Sullivan, Hirohisa Tanaka, Benjamin Quilain and Michael Wilking for assistance using the WCSim and FitQun software. We would like to also thank Marie-Laure Schneider for her help in preparing this article.